\documentclass[12pt]{article} 
\usepackage{amsmath,amssymb} 
\usepackage{epsfig}
\begin{document} 
\setlength{\parskip}{0.45cm} 
\setlength{\baselineskip}{0.75cm} 
%
%
%
\begin{titlepage} 
\setlength{\parskip}{0.25cm} 
\setlength{\baselineskip}{0.25cm} 
\begin{flushright} 
DO-TH 2001/04\\ 
\vspace{0.2cm} 
hep--ph/0103216\\ 
\vspace{0.2cm} 
March 2001 
\end{flushright} 
\vspace{1.0cm} 
\begin{center} 
\LARGE 
{\bf The Photon Structure Function at Small-x} 
\vspace{1.5cm} 
 
\large 
M. Gl\"uck, E.\ Reya, I.\ Schienbein\\ 
\vspace{1.0cm} 
 
\normalsize 
{\it Universit\"{a}t Dortmund, Institut f\"{u}r Physik,}\\ 
{\it D-44221 Dortmund, Germany} \\ 
\vspace{0.5cm}

\vspace{1.5cm} 
\end{center} 
 
\begin{abstract} 
\noindent 
It is shown that recent small--$x$ measurements of the photon 
structure function $F_2^{\gamma}(x,Q^2)$ by the LEP--OPAL 
collaboration are consistent with parameter--free QCD predictions 
at all presently accessible values of $Q^2$. 
\end{abstract} 
\end{titlepage} 
 
 
Recently the OPAL collaboration \cite{ref1} at the CERN--LEP 
collider has extended the measurements of the photon structure 
function $F_2^{\gamma}(x,Q^2)$ into the small--$x$ region down 
to $x\simeq 10^{-3}$, probing lower values of $x$ than ever 
before.  The observed rise of $F_2^{\gamma}$ towards low values 
of $x$, $x<0.1$, is in agreement with general QCD renormalization 
group (RG) improved expectations.  It has, however, been noted 
that the rising small--$x$  data at lower scales $Q^2\simeq 2-4$ 
GeV$^2$ lie above the original QCD expectations anticipated 
almost a decade ago \cite{ref2,ref3}. 
 
It is the purpose of the present note to demonstrate that more 
recent and updated parameter--{\underline{free}} QCD predictions \cite{ref4} 
for $F_2^{\gamma}(x,Q^2)$ are in general also consistent with 
\mbox{the OPAL} small--$x$ measurements at all presently accessible values 
of $Q^2$. 
 
Before presenting our results it is instructive to recapitulate 
briefly the main differences between the original GRV$_{\gamma}$ 
\cite{ref2} approach to the photonic parton distributions and 
the more recent parameter--free predictions of GRS \cite{ref4}. 
In the latter approach a coherent superposition of vector mesons 
has been employed, which maximally enhances the $u$--quark  
contributions to $F_2^{\gamma}$, for determining the hadronic 
parton input $f_{\rm had}^{\gamma}(x,Q_0^2)$ at a GRV--like 
\cite{ref5} input scale $Q_0^2\equiv \mu_{\rm LO}^2=0.26$ GeV$^2$ 
and $Q_0^2\equiv \mu_{\rm NLO}^2=0.40$ GeV$^2$ for calculating 
the (anti)quark and gluon distributions $f^{\gamma}(x,Q^2)$ of 
a real photon in leading order (LO) and next--to--LO (NLO) of 
QCD.  Furthermore, in order to remove the ambiguity of the  
hadronic light quark sea and gluon input distributions of the 
photon (being related to the ones of the pion, $f^{\pi}(x,Q_0^2)$, 
via vector meson dominance), inherent to the older GRV$_{\gamma}$ 
\cite{ref2} and SaS \cite{ref3} parametrizations, predictions 
\cite{ref6} for $\bar{q}\,^{\pi}(x,Q^2)$ and $g^{\pi}(x,Q^2)$ 
have been used by GRS \cite{ref4} which follow from constituent 
quark model constraints \cite{ref7}.  These latter constraints 
allow to express $\bar{q}\,^{\pi}$ and $g^{\pi}$ entirely in terms 
of the experimentally known pionic valence density and the 
rather well known quark--sea and gluon distributions of the  
nucleon \cite{ref6}, using most recent updated valence--like 
input parton densities of the nucleon.  Since more recent  
DIS small--$x$ measurements at HERA imply somewhat less steep 
sea and gluon distributions of the proton \cite{ref5}, the  
structure functions of the photon will therefore also rise less 
steeply in $x$ \cite{ref4} than the previous GRV$_{\gamma}$ 
\cite{ref2} ones as will be seen in the figures shown below. 
In this way one arrives at truly parameter--free predictions 
for the structure functions and parton distributions of the  
photon. 
 
In Figs.\ 1 and 2 we compare the more recent GRS predictions 
\cite{ref4} and the older GRV$_{\gamma}$ results \cite{ref2} 
with the recent small--$x$ OPAL measurements \cite{ref1} and, 
for completeness, some relevant L3 data \cite{ref8} are shown 
as well.  The parameter--free LO-- and NLO--GRS expectations  
are confirmed by the small--$x$ OPAL data at {\underline{all}} 
(small and large) experimentally accessible scales $Q^2$. 
This is in contrast to the GRV$_{\gamma}$ and SaS results  
which at LO are somewhat below the data at small $Q^2$ in Fig.\ 1 
and seem to increase too strongly at small $x$ in NLO, in  
particular at larger values of $Q^2$ as shown in Fig.\ 2.   
The main reason for this latter stronger and steeper $x$--dependence 
in LO and NLO derives from the assumed vanishing (pionic) quark--sea 
input at $Q_0^2=\mu_{\rm LO,NLO}^2$ for the anti(quark)  
distributions of the photon as well as from relating the 
hadronic gluon input of the photon directly to its (pionic) 
valence distribution \cite{ref2,ref9}.  This is in contrast 
to the more realistic (input) boundary conditions employed by 
GRS \cite{ref4,ref6}. 
 
Clearly these small--$x$ measurements imply that the photon must 
contain \cite{ref1} a dominant hadron--like component at low 
$x$, since the simple direct `box' cross section (based on the 
subprocess $\gamma^*(Q^2)\gamma\to q\bar{q}\,$) yields  
$F_{2,{\rm box}}^{\gamma}\to 0$ as $x\to 0$, in contrast to the 
data for $x<0.1$ in Figs.\ 1 and 2.  The QCD RG--improved parton 
distributions of the photon are thus essential for understanding 
the data on $F_2^{\gamma}(x,Q^2)$, with its dominant contributions 
deriving from $q^{\gamma}(x,Q^2)=\bar{q}\,^{\gamma}(x,Q^2)$. 
It would be also interesting and important to extend present 
measurements \cite{ref10,ref11} of the gluon distribution of 
the photon, $g^{\gamma}(x,Q^2)$, {\underline{below}} the presently 
measured region  
0.1 \raisebox{-0.1cm}{$\stackrel{<}{\sim}$} $x<1$  
where similarly $g^{\gamma}(x<0.1,\,Q^2)$ is expected to be also 
somewhat flatter \cite{ref4} in the small--$x$ region than 
previously anticipated \cite{ref2}. 
 
%
 
\newpage

\noindent{\large{\bf{\underline{Figure Captions}}}} 
\begin{itemize} 
\item[\bf{Fig.\ 1}.] Comparison of the parameter--free GRS  
      predictions \cite{ref4}, the previous GRV$_{\gamma}$ 
      \cite{ref2} and SaS \cite{ref3} results for $F_2^{\gamma} 
      (x,Q^2)$ with the recent OPAL (1.9 GeV$^2$) small--$x$ measurements 
      \cite{ref1} at two fixed lower scales $Q^2$.  The previous 
      OPAL (1.86 GeV$^2$) \cite{ref1} and L3 \cite{ref8} data are also 
      displayed. 
       
\item[\bf{Fig.\ 2}.]  As in Fig.\ 1 but at two fixed scales 
      $Q^2$.  The recent OPAL small--$x$  data are taken from 
      Ref.\ \cite{ref1}. 
\end{itemize} 

\newpage
\pagestyle{empty}
\begin{figure}
\centering
\vspace*{-1cm}
\epsfig{figure=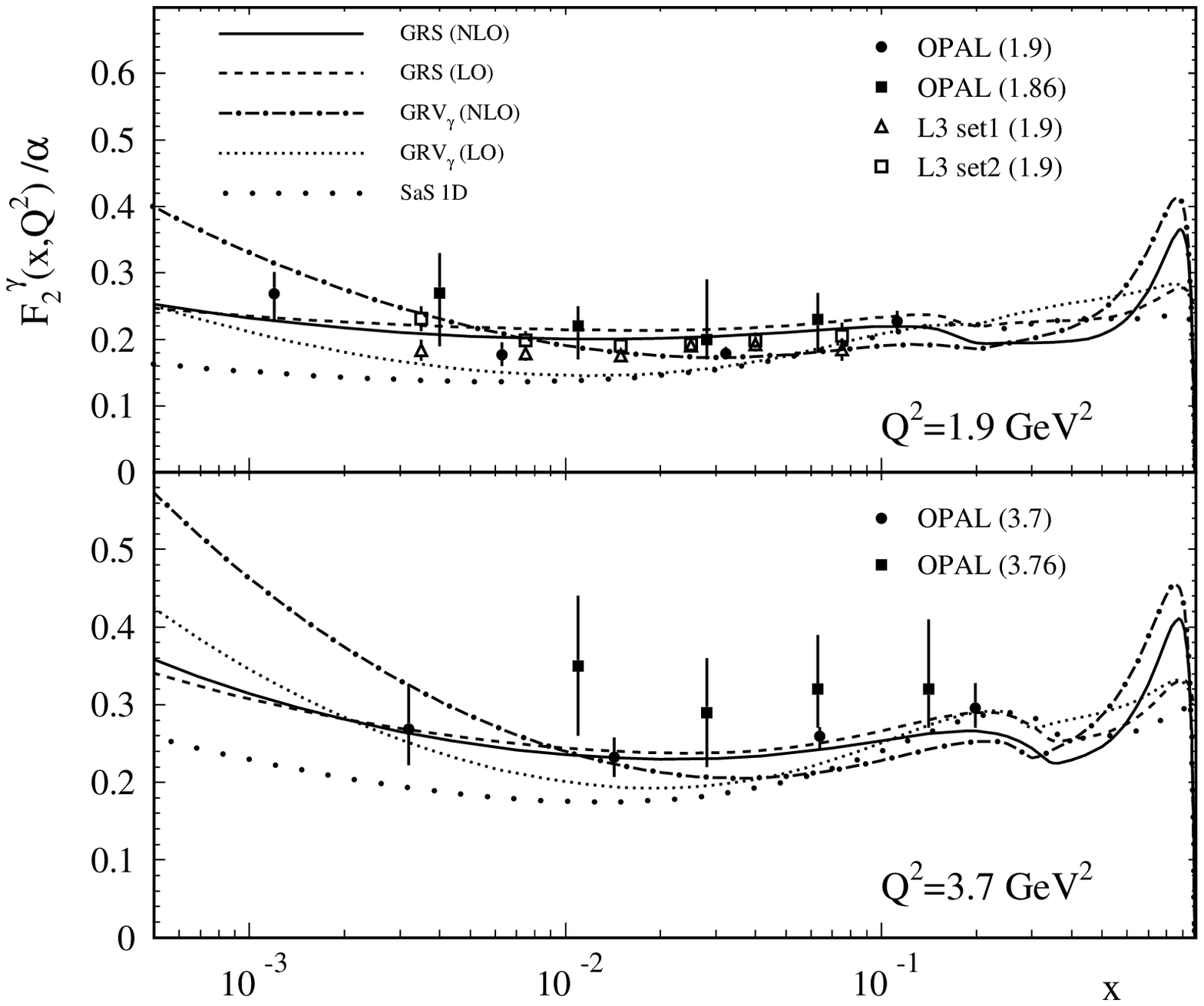,width=16cm}

\vspace*{0.3cm}
{\large\bf Fig. 1}
\end{figure}

\newpage
\pagestyle{empty}
\begin{figure}
\centering
\vspace*{-1cm}
\epsfig{figure=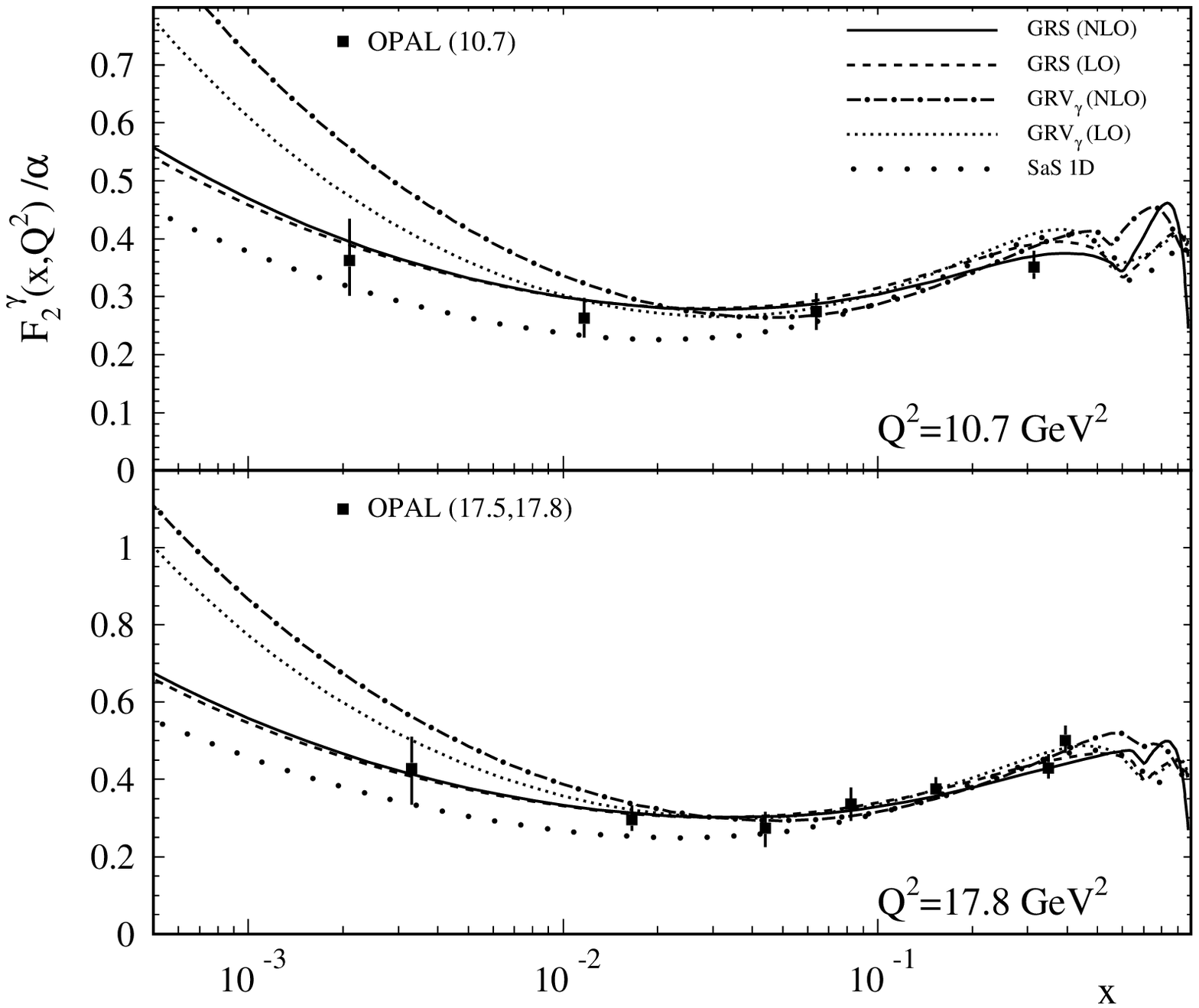,width=16cm}

\vspace*{0.3cm}
{\large\bf Fig. 2}
\end{figure}

\end{document}